\documentclass[%
 reprint,
 amsmath,amssymb,
 aps,
]{revtex4-2}

\usepackage[T1]{fontenc} 
\usepackage{graphicx}
\usepackage{float}
\usepackage{dcolumn}
\usepackage{bm}

\begin{document}

\preprint{Physical Review Letters}

\title{Decoding the Density Dependence of the Nuclear Symmetry Energy}

\author{W. G. Lynch}
    \email{lynch@nscl.msu.edu}
\author{M. B. Tsang}
    \email{tsang@nscl.msu.edu}
\affiliation{
    National Superconducting Cyclotron Laboratory, East Lansing, MI, 48824, USA.\\
    Department of Physics and Astronomy, Michigan State University, East Lansing, MI 48824, USA.
}
\date{\today}

\begin{abstract}
    The large imbalance in the neutron and proton densities in very neutron rich systems increases the nuclear symmetry energy so that it governs many aspects of neutron stars and their mergers. Extracting the density dependence of the symmetry energy therefore constitutes an important scientific objective. Many analyses have been limited to extracting values for the symmetry energy, $S_0$, and its ``derivative'', $L$, at saturation density $\rho_0 \approx 2.6 \times 10^{14}~\mathrm{g/cm^3}$ $\approx 0.16~\mathrm{nucleons/fm^{3}}$, resulting in constraints that appear contradictory. We show that most experimental observables actually probe the symmetry energy at densities far from $\rho_0$, making the extracted values of $S_0$ or $L$ imprecise. By focusing on the densities these observables actually probe, we obtain a detailed picture of the density dependence of the symmetry energy from $0.25\rho_0$ to $1.5\rho_0$. From this experimentally derived density functional, we extract $L_{01}=53.1\pm6.1 MeV$ at $\rho \approx 0.10~\mathrm{fm^{-3}}$, a neutron skin thickness for $^{208}Pb$ of $R_{np} =$ $0.23\pm0.04$ fm, a symmetry pressure at saturation density of $P_0=3.2\pm1.2 MeV/fm^3$ and suggests a radius for a 1.4 solar mass neutron star of $13.1\pm0.6$ km.  
\end{abstract}

\maketitle



The nuclear Equation of State (EOS) is central to describing matter within neutron stars (NS) and explosive stellar environments, including NS mergers and core collapse supernovae ~\cite{lattimer2004physics, lattimer2016equation, steiner2009constraints, roberts2012medium, abbott2017gw170817,abbott2018gw170817}. To describe such neutron-rich environments one must extrapolate the properties of neutron-rich matter from that of symmetric matter with equal neutron and proton densities ~\cite{danielewicz2002determination, horowitz2014way, li2008recent}. This extrapolation is governed by the nuclear symmetry energy (SE), which can be defined as the difference between the EOS of pure neutron matter and that of symmetric matter. 

While many investigations have focused on the expansion of the symmetry energy near saturation density, $\rho_0$, 
the ultimate goal is to obtain the SE functional over a wide range of densities, from well below to far above saturation density. Knowledge of the SE at $\rho/\rho_0 \approx 0.25$ is important for the understanding of the supernova neutrino sphere ~\cite{roberts2012medium, horowitz2006cluster, martinez2012charged} and at $0.5 < \rho/\rho_0 < 0.7$ to predict the crust-core boundary ~\cite{hebeler2013equation} and crustal vibrations ~\cite{steiner2010equation} in NS. The region with $\rho > \rho_0$ is essential to determine the radii and tidal deformabilites of NS and the proton fractions within them. 

In this letter, we show how to combine results from different observables ~\cite{tsang2009constraints,tsang2012constraints, lattimer2013constraining, li2013constraining,kortelainen2010nuclear, danielewicz2014symmetry, roca2015neutron, tamii2011complete, brown2013constraints} to constrain the SE at densities from 0.25 to 0.7 $\rho_0$.  
Then we include recent measurements of pion spectral ratios and the neutron skin of $^{208}Pb$, $R_{np}$, and extend the density functional to suprasaturation densities approaching 1.5 $\rho_0$.   

For the asymmetries achieved experimentally, the SE can be approximated by $\varepsilon_\mathrm{sym} = S(\rho)\delta^2$. Here, $S(\rho)$ describes the density dependence of the SE, $\delta = (\rho_n - \rho_p) / \rho$ is the isospin asymmetry, and $\rho_n$, $\rho_p$ and $\rho = \rho_n + \rho_p$ are the neutron, proton and total densities, respectively ~\cite{li2008recent}. $S(\rho)$, is frequently discussed in terms of an expansion around saturation density, $\rho_0$, 
\begin{equation}
    S(\rho) = S_0 + \frac{L}{3\rho_0}(\rho - \rho_0) + \frac{K_\mathrm{sym}}{18\rho_0^2}(\rho - \rho_0)^2 + \cdots,
\end{equation}
where $L = 3\rho_0\left.\frac{S(\rho)}{\partial \rho}\right\rvert_{\rho=\rho_0}$ and $K_\mathrm{sym}=9\rho_0^{2}\left.\frac{\partial^{2} S(\rho)}{\partial \rho^{2}}\right\rvert_{\rho=\rho_0}$. 
 
At saturation density, the symmetry pressure $P_0 = (L\rho_0)/3$ governs the pressure of neutron matter ~\cite{li2008recent}. Constraints on $S_0$ and $L$ have been obtained with a variety of nuclear observables, but the uncertainties in  $S_0$ and $L$ remain significant~\cite{horowitz2014way, li2008recent, tsang2009constraints,tsang2012constraints,lattimer2013constraining, li2013constraining, kortelainen2010nuclear, danielewicz2014symmetry, roca2015neutron, tamii2011complete, brown2013constraints}. 
Attempts to reduce the uncertainties in $S_0$ and $L$ by averaging  values from different analyses  ~\cite{li2013constraining} or by considering the overlaps of the various $S_0$ and $L$ contours ~\cite{lattimer2013constraining} does not provide the SE density functional. In the following, we decode published constraints for five observables to find the ``sensitive density'' $\rho_s$ at which each observable accurately provides $S(\rho_s)$. Typically, $\rho_s$ is far from $\rho_0$. 
We then combine these $S(\rho_s)$ values with additional measurements to constrain the SE density functional $S(\rho)$ at $0.25 < \rho/\rho_{0}<1.5.$

\begin{figure}
    \centering
    \includegraphics[width=\columnwidth]{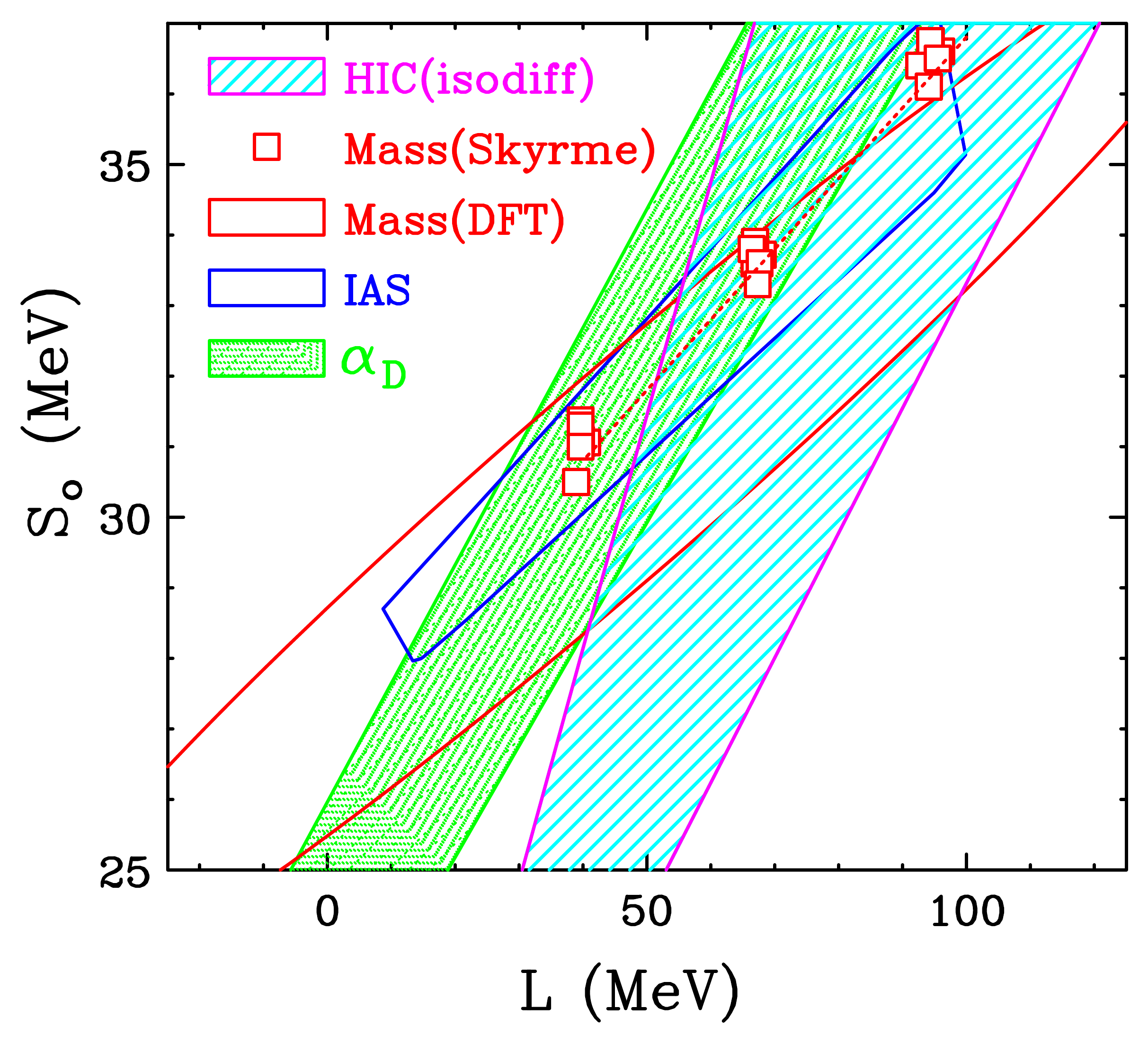}
    \caption{
        (color online): Contours for the allowed values of $(S_0, L)$ obtained from analyses of diffusion, (HIC(isodiff)), masses (Mass(DFT)), isobaric analog states (IAS), and electric dipole polarizability ($\alpha_\mathrm{D}$) of $^{208}\mathrm{Pb}$  ~\cite{horowitz2014way}. The open squares indicate the $S_0$ and $L$ values for 18 SE functionals obtained by fits of doubly closed shell nuclei in Ref. ~\cite{brown2013constraints}. The dotted line is the best fit of the open square points.
    }
\end{figure}

The bounded regions in Fig. 1 illustrate the nature of the correlated contours on $S_0$ and $L$ from Ref. ~\cite{horowitz2014way} that have been extracted from the analysis of four different structure and reaction observables. For clarity, this figure shows only one constraint region for each observable, even when others exist ~\cite{lattimer2013constraining}. The four bounded regions in Fig. 1 correspond to: 1) Mass(DFT): analyses of nuclear masses (red ellipse)~\cite{kortelainen2010nuclear, kortelainen2012nuclear}, 2) IAS: analyses of isobaric analog states (blue contour) ~\cite{danielewicz2014symmetry}, 3) HIC(isodiff): analyses of isospin diffusion in Sn+Sn collisions (blue shaded region)~\cite{tsang2009constraints, tsang2004isospin, liu2007isospin} and 4) $\alpha_D$: analyses of the electric dipole polarizability of $^{208}\mathrm{Pb}$ (green shaded region)~\cite
{roca2015neutron, tamii2011complete, roca2013electric}. To these four contours, we add 5) Mass(Skyrme): a set of  the correlated $S_0$ and $L$ values (shown as 3 groups of open squares) from analyses of the masses of doubly magic nuclei ~\cite{brown2013constraints} . 

Each constraint has a long ``principal'' axis with a different inclination, $\tau=\Delta S_0/\Delta L$\footnote{For clarity, we use the term inclination to describe the direction of the principal axis instead of ``slope'' to avoid confusing it with $L$, also known as slope at saturation density}. While the $\tau$ values (listed in Table I) from Mass(Skyrme) \cite{brown2013constraints}, Mass(DFT) \cite{kortelainen2012nuclear} and IAS contours \cite{danielewicz2017symmetry} are similar, those obtained from $\alpha_D$ \cite{tamii2011complete,tamii2014electric,roca2013electric,roca2015neutron} and HIC(isodiff) \cite{tsang2004isospin,tsang2009constraints,liu2007isospin} are much steeper. We will show how the inclination reflects the density probed by the corresponding observable.

Values of $S_0$ and $L$ that describe an observable equally well can be found on contours of constant probability near the center of the principal axis. In this ``best fit'' region, two neighboring points ($L,S_{0})_i$ and ($L,S_{0})_j$ on a contour can simultaneously define the direction $\tau$=$\Delta$$S_{0}$/$\Delta$$L$ of the principal axis and describe how values for $S_{0}$  can be compensated by small changes in $L$ so  that the modified $S(\rho)$ will still reproduce the experimental data equivalently.

Some observables primarily probe $S(\rho)$ within a narrow density range. Retaining a good fit requires $S(\rho)$ to remain invariant at $\rho$ while $S_o$ and $L$ are varied.This connects the $\tau$ value describing these variations to $S(\rho)$ as follows:

\begin{equation}
    \tau = \frac{\Delta S_0}{\Delta L} = -\left.\frac{\partial S(\rho)}{\partial L} \middle/ \frac{\partial S(\rho)}{\partial S_0}\right. \ .
\end{equation}

\begin{figure}[ht]
    \centering
    \includegraphics[width=\columnwidth]{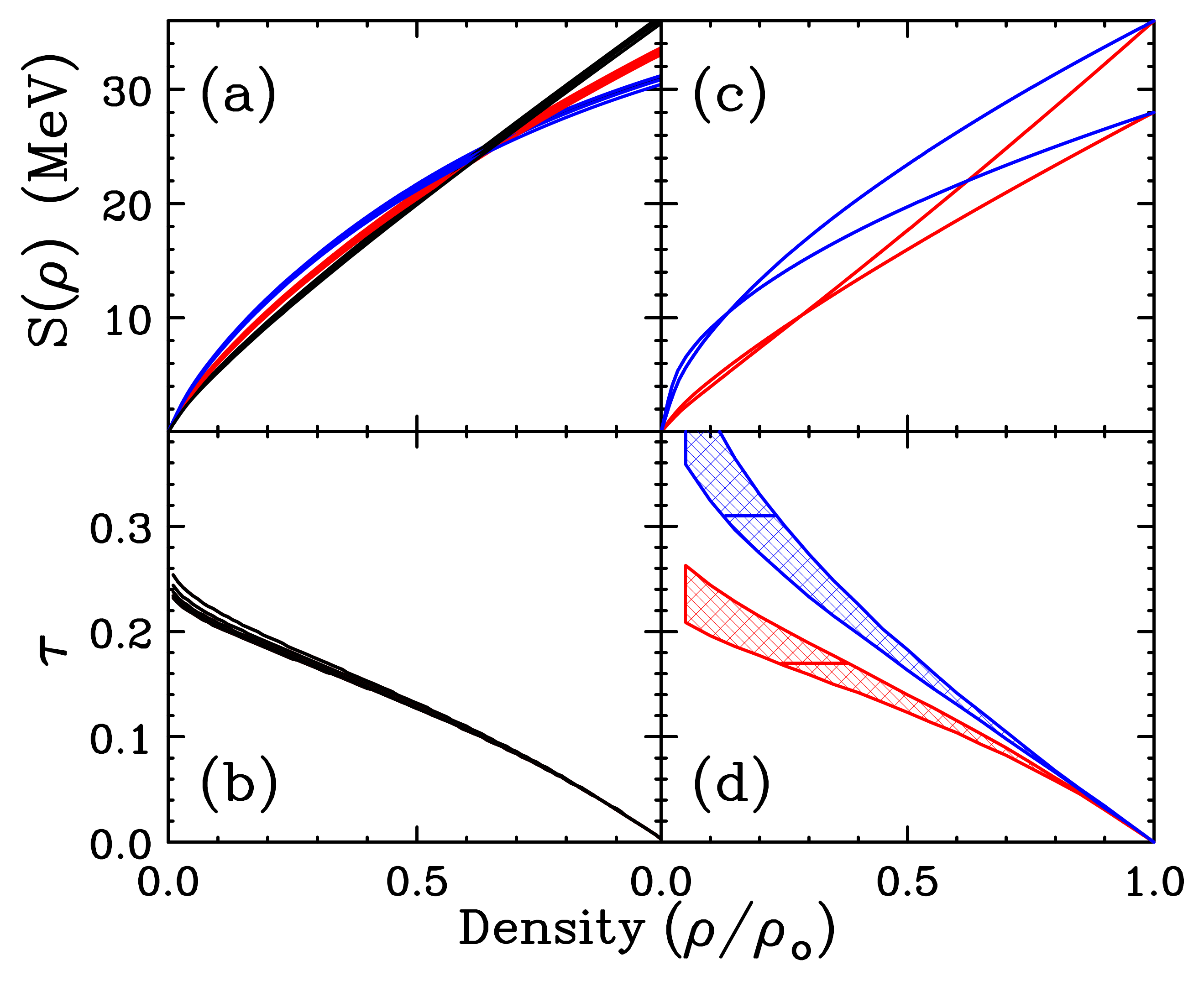}
    \caption{(a) 18 Skyrme SE functionals of Eq. (3) that reproduce masses of doubly magic nuclei ~\cite{brown2013constraints} 
    (b) Density dependence of $\tau$ in Eq. (3), (c) Two parameter Skyrme density functionals from Eq. (4). The top (blue) group of calculations corresponds to  left boundary of HIC(isodiff) contours while the lower (red) group of calculations  represents the right boundary of the contour. (d) Density dependence of $\tau$ of Eq. (4) corresponding to the upper and lower boundaries of the iso(diff) constraints in Fig. 1.}
\end{figure}

\noindent To illustrate how $\tau$  connects the SE to a specific sensitive density $\rho_s$, we consider the work of Brown ~\cite{brown2013constraints}, who fitted the energies of 11 doubly closed shell nuclei with Skyrme interaction in the form of: 
\begin{equation}
    S(\rho) = a\left(\frac{\rho}{\rho_0}\right) + b\left(\frac{\rho}{\rho_0}\right)^{1+\sigma} + c\left(\frac{\rho}{\rho_0}\right)^{\frac{2}{3}} + d\left(\frac{\rho}{\rho_0}\right)^{\frac{5}{3}}.
\end{equation}
This functional form is often used to calculate masses and other nuclear properties using Hartree-Fock density functional theory ~\cite{brown2013constraints, dutra2012skyrme}. Parameters $a$, $b$, $c$, $d$ and $\sigma$ fully define $S_0$ and $L$.

Fig. 2(a) shows 18 SE functionals used by Brown ~\cite{brown2013constraints}. The three groups of calculations that provide $R_{np} = 0.16$, $0.20$ and $0.24~\mathrm{fm}$, with $L \approx 40$, $60$ and $90~\mathrm{MeV}$, respectively~\cite{brown2013constraints, brown2014constraints} converge (cross-over) to  $S(\rho_s)= 24.7\pm0.8\mathrm{MeV}$ at $\rho_s/\rho_0 = 0.63\pm0.03$. The SE functionals with a larger (smaller) $L$ have a larger (smaller) skin and a stronger (weaker) density dependence of $S(\rho)$. The corresponding $S_0$ vs $L$ values for these 18 calculations, shown as the open squares in Fig. 1, also separate into 3 groups. The dotted line provides the linear best fit through the points and a value of $\tau$ = $0.100\pm0.006$.

Fig. 2(b) shows the calculated density dependence of $\tau$ for the 18 SE functionals 
These functions attain $\tau$ = $0.100\pm0.006$ only at $\rho_s/\rho_{0}=0.63\pm0.03$. S($\rho_s$) and its uncertainty can then be determined by inserting $\rho_s$ into the 18 "best fit" SE functionals.  In this case, the ``Cross-over'' and the ``Inclination'' analyses determine the same values for $\rho_s$ and for $S(\rho_s)$. 

We next examine constraints from the isospin diffusion data which has much steeper $\tau$ values. These were modeled with the two parameter SE density functional in Eq. 4, which has been widely employed in transport model calculations of nuclear collisions~\cite{li2008recent, zhang2008influence, hong2014subthreshold}: 
\begin{equation}
    S(\rho) = S_\mathrm{kin}+S_\mathrm{int}=A\left(\frac{\rho}{\rho_0}\right)^{\frac{2}{3}} + B\left(\frac{\rho}{\rho_0}\right)^\gamma.
\end{equation}
Here, $S_\mathrm{kin}$ and $S_\mathrm{int}$ model the kinetic and interaction contributions to the SE, respectively. 

Isospin diffusion is observed in peripheral collisions involving collisions of nuclei with different isospin asymmetries. The isospin diffusion rate between projectile and target is primarily controlled by the magnitude of the SE of the low density matter that connects them during the collision  ~\cite{shi2003nuclear}. The region labeled HIC(isodiff) is obtained by modeling the isospin diffusion data of $^{124,112}\mathrm{Sn}+^{112,124}\mathrm{Sn}$ at E/A=50 MeV in Ref. ~\cite{tsang2009constraints} with  Eq. (4) and the transport model code ImQMD05  ~\cite{zhang2008influence}.

Fig. 2(c) shows that the SE functionals used in Ref.~\cite{tsang2009constraints} intersect at low densities where the SE approaches 10.4 MeV. 
The figure focuses on the two crossing points that occur for $(S_{0},L)$ values on the boundaries of the HIC(isodiff) in Fig. 1, where the agreement between the data and the transport calculations is at the 2$\sigma$ level. 

These crossing points indicate the sensitive densities probed by isospin diffusion. The left (low $L$) boundary with steeper inclination crosses-over  at $\rho_s/\rho_0=0.15$. The right (high $L$) boundary  crosses-over at a higher density of $\rho_s/\rho_0=0.29$. Together these two crossing points combine to provide the values of $S(\rho_s)= 10.3\pm1.0\mathrm{MeV}$ at $\rho_s/\rho_0 = 0.22\pm0.07$.

Similar constraints can be derived from the inclination of the HIC(isodiff) boundaries. Fig. 2(d) shows the corresponding density dependence of $\tau$ for the two sets of functionals shown in Fig. 2(c). 
Unlike Fig. 2(b), the density dependence of $\tau$ of Eq. (4) do not collapse onto a universal curve but span an area represented by the shaded region. 

Indeed, the horizontal line segments in the cross-hatched regions in Fig. 2(d) represent the $\tau$ values given by the high $L$ (red) and the low $L$ (blue) boundaries in Fig. 1. From the intersection of these line segments with the boundaries, we obtain the maximum and minimum densities $\rho_s/\rho_0 = 0.21\pm0.11$ with $S(\rho_s)= 10.1\pm1.0\mathrm{MeV}$.

The results from both the Inclination and Cross-over methods are consistent within uncertainties. Due to its higher precision, we plot the cross-over values by the open star in Fig. 3, where it constrains the EoS at the low density regions relevant to the neutrino-sphere of a core-collapse supernova ~\cite{roberts2012medium}.

\begin{table*}
    \begin{ruledtabular}
    \begin{tabular}{ccccccc}
        &  & \multicolumn{2}{c}{Inclination analyses} & \multicolumn{2}{c}{Cross-over analyses}& Ref. \\
        Constraint & $\tau$ & $\rho_s/\rho_0$ & $S(\rho_s)$ (MeV) & $\rho_s/\rho_0$ & $S(\rho_s)$ (MeV)& \\
        \hline
        Mass(Skyrme) & $0.100\pm0.006$ & $0.63\pm0.03$ & $24.7\pm0.8$ & $0.63\pm0.03$ & $24.7\pm0.8$ &~\cite{brown2013constraints}\\
        Mass(DFT) & $0.079\pm0.002$ & $0.72\pm0.01$ & $25.4\pm1.1$& & & ~\cite{kortelainen2012nuclear} \\
        IAS & $0.092\pm0.008$ & $0.66\pm0.04$ & $25.5\pm1.1$ & & & ~\cite{danielewicz2017symmetry}\\
        HIC(isodiff)  & $0.256\pm0.076$ & $0.21\pm0.11$ & $10.1\pm1.0$ & $0.22\pm0.07$ & $10.4\pm1.0$&~\cite{tsang2009constraints} \\
        \hline
        &  & & & \multicolumn{2}{c}{From publications} &  \\
            \hline
         $\alpha_D$  &   & &   &   $0.31\pm0.03$ & $15.9\pm1.0$&~\cite{zhang2015electric}\\
         HIC(n/p) &   &   &  & $0.43\pm0.05$ & $16.8\pm1.2$&~\cite{morfouace2019constraining} \\
          HIC($\pi$) &   &  &   &  $1.5\pm0.2$ & $52\pm13$&~\cite{estee2021probing} \\
        PREXII ($^{208}\mathrm{Pb}$ skin) &   &  &   &  $1\pm0$ & $38.1\pm4.7$&~\cite{adhikari2021accurate} \\      
    \end{tabular}
    \end{ruledtabular}
    \caption{Values of $\tau$, $\rho_s$ and $S(\rho_s)$, obtained from the direct examination of the SE cross-over points and from the inclination of the ($S_0, L$) correlation for different experimental observables in Fig. 1 (upper half). For HIC(isodiff), there are two boundaries with different inclinations. The $\tau$ values listed span both boundaries. Lower half of the table lists constraints obtained from publications.}
\end{table*}

The IAS and DFT contours in Fig. 1 use the functional form of  Eq. (3). Since their $\tau$ values listed in Table I are very similar to that of Mass(Skyrme), one can use Fig. 2(b) to provide the $\rho_s$ and $S(\rho_s)$ values with little ambiguity. The results are plotted as a solid blue circle and open blue triangle in Fig. 3, respectively. The sensitive densities for these constraints are slightly different, reflecting differences in the masses of the nuclei fitted. The DFT partial ellipse in Fig. 1 fitted masses of nuclei with $40\leq A \leq 264$ with a greater emphasis on heavier nuclei, while the IAS polygon fitted Isobaric Analog States with $30 \le A \le 240$ with a lesser emphasis on heavy nuclei. 

\begin{figure}
    \centering
    \includegraphics[width=\columnwidth]{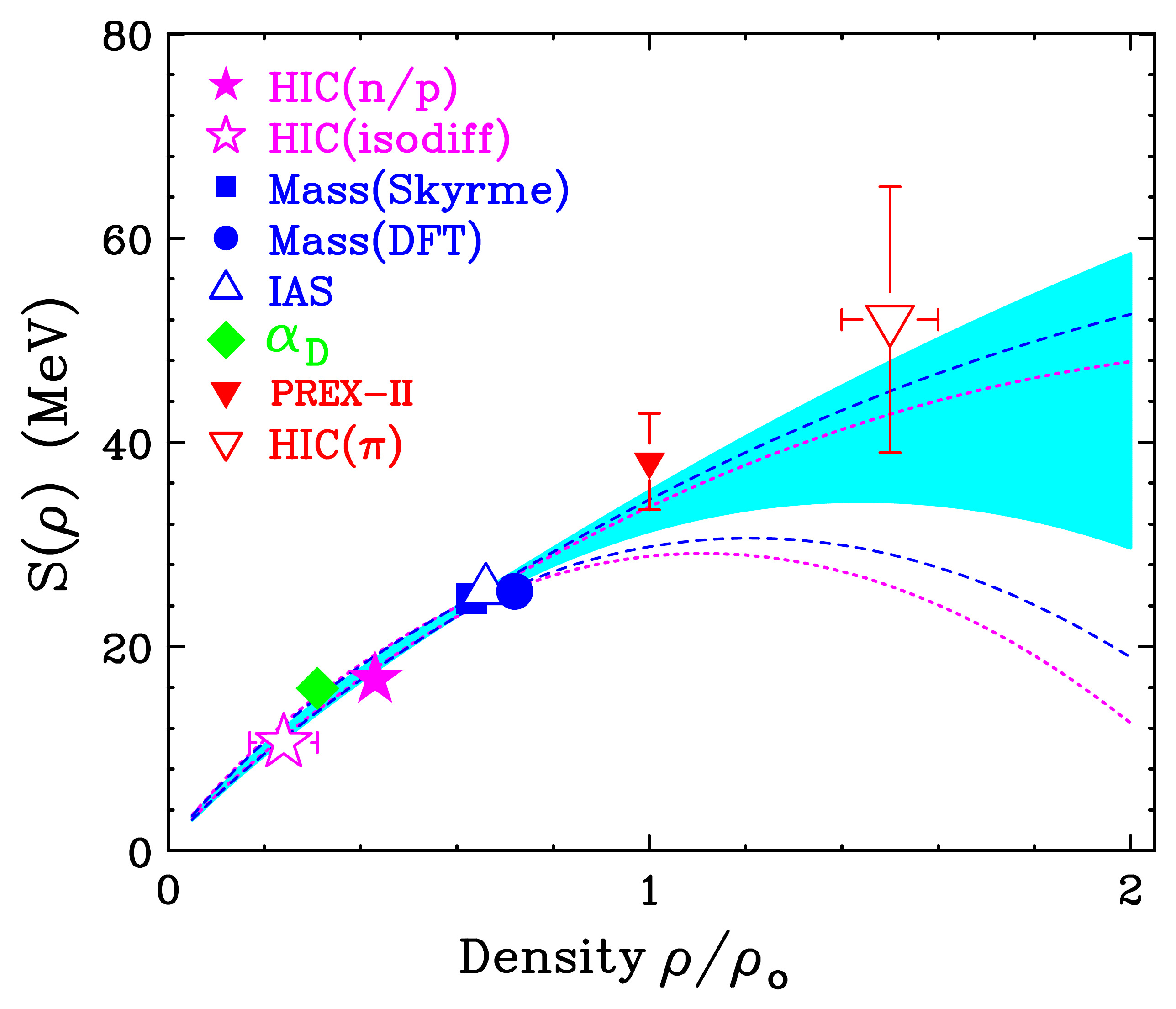}
    \caption{Density dependence of the symmetry energy. Symbols are data discussed in the text. The shaded area show the $1\sigma$ contours from fitting Eq. 6. See text for details.   The dotted curves represent upper and lower bounds of fits without PREXII and pion results while the dashed curves include PREXII but no pion results. }
\end{figure}

 To these data points we now add complimentary constraints from other observables obtained via Bayesian and Pearson correlation analyses ~\cite{zhang2013constraining}. 
First we include the dipole polarizability contour $\alpha_\mathrm{D}$, which reflects the response of the  $^{208}Pb$ nucleus to the presence of an external electric field. Pearson correlation analysis of Zhang et al. ~\cite{zhang2015electric} indicates that $\alpha_\mathrm{D}$ constrains the SE most accurately at $\rho_s/\rho_0 \approx 0.31$, with $S(\rho_s) = 15.9\pm1.0~\mathrm{MeV}$. This point is plotted as the solid green diamond in Fig. 3 and is listed in the lower half of Table I. The upper half of the table contains results using the Cross-over and Inclination analyses as discussed above. 

 

In Ref. \cite{morfouace2019constraining} a cross-over was observed in the SE functionals that best described the single and double ratios of neutron and proton energy spectra for central $^{124}Sn+^{124}Sn$ and $^{112}Sn+^{112}Sn$ collisions at $E/A = 120~\mathrm{MeV}$ ~\cite{morfouace2019constraining}. Bayesian analyses of these data result in the HIC(n/p)  data point listed in table I and plotted as a solid star in Fig. 3 \footnote{One of the figures in Ref. ~\cite{morfouace2019constraining} includes most of the low density data points of Fig 3. They were taken from an earlier unpublished version of the present work that it cites ~\cite{lynch2018nuclear}}. 

We now include recent SE constraints at higher densities. Pion emission in heavy ion reactions has recently yielded a SE value listed in Table I ~\cite{estee2021probing} at $\rho \approx$ 1.5 $\rho_0$ ~\cite{liu2021insights}.  The open inverted triangle in Fig. 3 shows this data point  obtained from the ratio of charged pion transverse momentum spectra for central collisions of $^{132}Sn+^{124}Sn$ and $^{108}Sn+^{112}Sn$ at $E/A=270~\mathrm{MeV}$ ~\cite{estee2021probing}. While further studies may more tightly constrain the most sensitive density and its uncertainty, the current data with its uncertainties suffice for the present work. 

An improved value of $R_{np}$=$(0.29\pm0.07) fm$ has also been obtained for the neutron skin thickness of $^{208}\mathrm{Pb}$ from the measurements of parity violating electron scattering~\cite{adhikari2021accurate}, referred as PREXII here. The skin thickness measurement provides a value for the first derivative of the SE functional at $\rho_{01}= 0.1 fm^{-3}$  corresponding to the Taylor expansion coefficient $L_{01}= 3\rho_{01}\left.\frac{S(\rho)}{\partial \rho}\right\rvert_{\rho_{01}}$~\cite{reed2021implications}. However, the experimental results are often reported in terms of  $L$ and $S_0$ values extrapolated to saturation density~\cite{reed2021implications}. We list these values in Table I and plot $S_0$ as the solid inverted triangle in Fig. 3.


We now fit the data with the expression: 
\begin{equation}
    S(\rho) = S_\mathrm{kin}(\rho)+S_\mathrm{int}(\rho)
\end{equation}
    where
\begin{equation}
    S_\mathrm{int}(\rho)=S_\mathrm{int}(\rho_{01})+S^{'}_\mathrm{int}(\rho-\rho_{01})+\frac{1}{2}S^{''}_\mathrm{int}(\rho-\rho_{01})^2.
\end{equation}
Here, $S_\mathrm{kin}=A(\rho/\rho_{0})^{2/3}$, $A=12.7 MeV$, and $S_\mathrm{int}$, $S^{'}_\mathrm{int}$, and $S^{''}_\mathrm{int}$ are parameterized so that $S_\mathrm{int}(0)=0$. 

The shaded region in Fig. 3 is the $1\sigma$ allowed region consistent with the best fit of experimental data points presented here and plotted in Fig. 3. For PREXII measurement, the fit includes the more accurate value  $L_{01}$=(71.46 ± 22.60) MeV rather than the extrapolated $L$ and $S_0$ values. (If the extrapolated $S_0$ and $L$ values are used instead of the $L_{01}$ values, the symmetry energy fit is slightly stiffer.) If we exclude both the pion and PREXII data, we obtain a much softer symmetry energy constraint with upper and lower fit boundaries represented by the dotted curves. The dashed curves enclose the best fit region with PREXII results included and the pion results excluded. While inclusion of the PREXII results stiffens the density dependence of the SE, adding the HIC($\pi$) data, even with its larger error bars, stiffens  the SE more significantly due its longer lever arm.  


With the SE density functional given by the shaded region in Fig. 3, we can extract quantities relevant to nuclei and to dense matter in neutron stars. Here we list the extracted values of $S_{01}$=$(24.2\pm0.5)~\mathrm{MeV}$, $L_{01}~\mathrm{MeV}$= $(53.1\pm6.1)~\mathrm{MeV}$ and $K_{01}$ =$(-79.2\pm37.6)~\mathrm{MeV}$. Since $L_{01}$ is tightly correlated with the $R_{np}$, our fit yields $R_{np}$= $(0.23\pm0.03)$ fm, within the experimental uncertainties of most $^{208}Pb$ skin measurements ~\cite{klimkiewicz2007nuclear, zenihiro2010neutron, wieland2009search, trzcinska2001neutron, roca2015neutron, tamii2011complete, pruitt2020systematic, atkinson2020dispersive} including the PREXII results ~\cite{reed2021implications, adhikari2021accurate}. 
The extracted NS radius of $13.1\pm0.6$ km is consistent with the results obtained in NICER ~\cite{miller2021radius, raaijmakers2021constraints, rini2021sizing} for both 1.4 and 2.08 solar mass neutron stars. From the compactness relation in Ref. ~\cite{tsang2020impact}, we obtain the deformability of $\Lambda = 500-720$ for 1.4 solar mass NS and $\Lambda =57-75 $ for 2.1 solar mass NS. The former value is near the high end of the revised GW170817 results ~\cite{abbott2017gw170817,abbott2018gw170817}. 

Our fit also directly provides $S_0$=$(33.3\pm1.3)~\mathrm{MeV}$, $L$= $(59.6\pm22.1)~\mathrm{MeV}$ and $K_\mathrm{sym}$ =$(-180\pm96)~\mathrm{MeV}$ and $P_0=3.2\pm1.2 MeV/fm^3$. From the correlation between $L$ and NS deformability in Ref. ~\cite{tsang2020impact}, we obtain $\Lambda$ values consistent with the above results. 




In summary, we decode the information contained in the disparate $S_0$ vs. $L$  contours used to represent the experimental symmetry energy constraints. We show  how these contours directly reflect the symmetry energy at the density favored by the experimental probe. We determine both the sensitive density and the corresponding symmetry energy for each measurement and combine these data to obtain a consistent description of the density dependence of the symmetry energy. 
Fitting the data obtained below the saturation density  favor a soft density dependence of the symmetry energy with a large uncertainties when extrapolating above $\rho_0$. Including the recent measurement of the PREXII results of $R_{np}$ stiffens the symmetry energy density dependence slightly but 
its influence is limited by the large experimental uncertainties. On the other hand, including the data at 1.5 $\rho_0$ from pion spectral ratios more effectively shifts the fit to favor stiffer SE density dependence even though the pion data have very large uncertainties. With the new symmetry energy functional, we should be able to calculate directly detailed properties of neutron stars including the crust-core transition pressure and the neutron star mass-radius relation, which will be the subject of a future publication.


\begin{acknowledgments}
The authors would like to thank Professor Hermann Wolter for giving us many insightful comments to our paper, Professor Alex Brown for providing the parameters of the 18 best fit Skyrme functions used in Ref. ~\cite{brown2013constraints}, Professor Pawel Danielewicz and Professor Witek Nazarewicz for providing some of the SE functions used to fit the IAS and DFT data,  Dr. Ingo Tews and Professor Jorge Piekarewicz for fruitful discussions. This work is supported by the US National Science Foundation Grant No. PHY-1565546, U.S. Department of Energy (Office of Science) under Grant Nos. DE-SC0014530, DE-NA0002923. 
\end{acknowledgments}


%

\end{document}